\documentclass[aps, prl, showpacs, twocolumn, superscriptaddress]{revtex4}

\usepackage{graphicx}
\usepackage{natbib}
\usepackage{dcolumn}
\usepackage{amssymb}

\begin{document}

\title{Generic Gravitational Wave Signals from the Collapse of
  Rotating Stellar Cores}

\author{H.~Dimmelmeier}
\affiliation{Department of Physics, Aristotle University of
  Thessaloniki, GR-54124 Thessaloniki, Greece}
\affiliation{Max-Planck-Institut f\"ur Astrophysik,
  Karl-Schwarzschild-Str.~1, D-85741 Garching, Germany}

\author{C.~D.~Ott}
\affiliation{Department of Astronomy and Steward Observatory,
  University of Arizona, Tucson, AZ 85721, USA}

\author{H.-T.~Janka}
\affiliation{Max-Planck-Institut f\"ur Astrophysik,
  Karl-Schwarzschild-Str.~1, D-85741 Garching, Germany}

\author{A.~Marek}
\affiliation{Max-Planck-Institut f\"ur Astrophysik,
  Karl-Schwarzschild-Str.~1, D-85741 Garching, Germany}

\author{E.~M\"uller}
\affiliation{Max-Planck-Institut f\"ur Astrophysik,
  Karl-Schwarzschild-Str.~1, D-85741 Garching, Germany}


\begin{abstract}
  We perform general relativistic (GR) simulations of stellar core
  collapse to a proto-neutron star, using a microphysical equation of
  state (EoS) and an approximation of deleptonization. We show that
  for a wide range of rotation rates and profiles the gravitational
  wave (GW) burst signals from the core bounce are generic, known as
  Type~I. In our systematic study, using both GR and Newtonian
  gravity, we identify and quantify the influence of rotation, the
  EoS, and deleptonization on this result. Such a generic type of
  signal templates will facilitate a more efficient search in current
  and future GW detectors of both interferometric and resonant type.
\end{abstract}

\pacs{04.25.Dm, 04.30.Db, 95.30.Sf, 95.30.Tg, 95.55.Ym, 97.60.Bw}
\maketitle


Theoretical predictions of the gravitational wave (GW) signal produced
by the collapse of a rotating stellar iron core to a proto-neutron
star (PNS) in a core collapse supernova are complicated, as the
emission mechanisms are very diverse. While the prospective GW burst
signal from the collapse, bounce, and early postbounce phase is
present only when the core rotates~\cite{mueller_82_a,
  moenchmeyer_91_a, zwerger_97_a, dimmelmeier_02_a, kotake_03_a,
  ott_04_a, ott_06_b}, one expects GW signals with sizeable amplitudes
also from convective motions at later post-bounce phases, anisotropic
neutrino emission, excitation of various oscillations in the PNS, or
nonaxisymmetric rotational instabilities~\cite{rampp_98_a,
  mueller_04_a, shibata_05_a, ott_06_a, ott_06_b}.

In the observational search for GWs from merging binary black holes,
powerful data analysis algorithms like matched filtering are applied,
as the waveform from the inspiral phase can be modeled very
accurately. In stark contrast, the GW burst signal from stellar
core collapse and bounce cannot yet be predicted with the desired
accuracy and robustness. First, a general relativistic (GR)
description of gravity and hydrodynamics including the important
microphysics is necessary. Only very few multi-dimensional codes have
recently begun to approach these requirements. Second, the rotation
rate and profile of the progenitor core are not very strongly
constrained. Therefore, the influence of rotation on the collapse
dynamics and thus the GW burst signal must be investigated by
computationally expensive parameter studies.

Previous simulations, considering a large variety of rotation rates
and profiles in the progenitor core but ignoring complex (though
essential) microphysics and/or the influence of GR, found
qualitatively and quantitatively different types of GW burst signals
(see, e.g., \cite{moenchmeyer_91_a, zwerger_97_a, dimmelmeier_02_a}).
These can be classified depending on the collapse dynamics:
\emph{Type~I} signals are emitted when the collapse of the
homologously contracting inner core is not strongly influenced by
rotation, but stopped by a \emph{pressure-dominated bounce} due to EoS
stiffening at nuclear density $ \rho_\mathrm{nuc} $ where the
adiabatic index $ \gamma_\mathrm{eos} $ rises above $ 4 / 3 $. This
leads to instantaneous PNS formation with a maximum core density
$ \rho_\mathrm{max} \ge \rho_\mathrm{nuc} $. \emph{Type~II} signals
occur when centrifugal forces grow sufficient to halt the collapse,
resulting in consecutive (typically multiple) \emph{centrifugal
  bounces} with intermediate coherent re-expansion of the inner core,
seen as density drops by often more than an order of magnitude; thus
here $ \rho_\mathrm{max} < \rho_\mathrm{nuc} $ after bounce.
\emph{Type~III} signals appear in a pressure-dominated bounce when the
inner core has a very small mass at bounce due to a soft subnuclear
EoS or very efficient electron capture.

In contrast, new GR simulations of rotational core collapse employing
a microphysical EoS and an approximation for deleptonization during
collapse suggest that the GW burst signature is exclusively of
Type~I~\cite{ott_06_b}. In this work we considerably extend the number
of models and comprehensively explore a wide parameter space of
initial rotation states. Also for this more general setup we find GW
signals solely of Type~I form. We identify the physical conditions
that lead to the emergence of this generic GW signal type and quantify
their relative influence. Our results strongly suggest that the
waveform of the GW \emph{burst} signal from the collapse of rotating
supernova cores is much more generic than previously thought.


We perform all simulations in 2\,+\,1 GR using the
\mbox{\textsc{CoCoNuT}} code~\cite{dimmelmeier_02_a,
  dimmelmeier_05_a}, approximating GR by the conformal flatness
condition (CFC), whose excellent quality in the context of rotational
stellar core collapse has been demonstrated extensively (see, e.g.,
\cite{shibata_04_a, ott_06_b}). \mbox{\textsc{CoCoNuT}} utilizes
spherical coordinates with the grid setup specified in~\cite{ott_06_b}
and assumes axisymmetry. GR hydrodynamics is implemented via
finite-volume methods, piecewise parabolic reconstruction, and an
approximate Riemann solver. We extract GWs using a variant of the
Newtonian quadrupole formula~(see, e.g., \cite{shibata_04_a}).

We employ the microphysical EoS of Shen et al.~\cite{shen_98_a,
  marek_05_a}. Deleptonization by electron capture onto nuclei and
free protons is realized as in~\cite{liebendoerfer_05_a}: During
collapse the electron fraction $ Y_e $ is parameterized as a function
of density based on data from spherically symmetric
neutrino-hydrodynamic simulations~\cite{marek_05_a} using the latest
available electron capture rates~\cite{langanke_00_a} (updating recent
results~\cite{ott_06_b} where standard capture rates were used). After
core bounce, $ Y_e $ is only passively advected and further lepton
loss is neglected. Contributions due to neutrino pressure $ P_\nu $
are taken into account above trapping
density~\cite{liebendoerfer_05_a}.

As initial data we take the non-rotating $ 20\,M_\odot $ progenitor
s20 from~\cite{woosley_02_a}, imposing the rotation law discussed
in~\cite{ott_04_a, dimmelmeier_02_a}. In order to determine the
influence of different angular momentum on the collapse dynamics, we
parameterize the initial rotation of our models in terms of the
differential rotation parameter $ A $ (A1: $ A = 50,000 \mathrm{\ km} $,
almost uniform; A2: $ A = 1,000 \mathrm{\ km} $, moderately
differential; A3: $ A = 500 \mathrm{\ km} $, strongly differential)
and the initial rotation rate $ \beta_\mathrm{i} = T / |W| $, which is
the ratio of rotational energy to gravitational energy (approximately
logarithmically spaced in 18 steps from $ 0.05\% $ to $ 4\% $).

\begin{figure}
  \includegraphics[width = 8.5 cm]{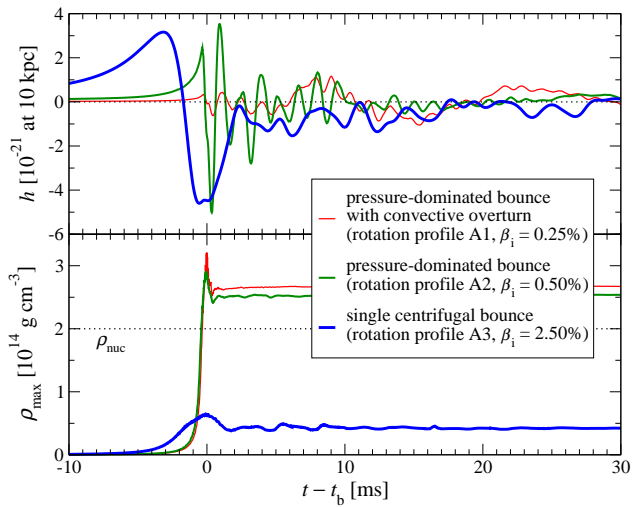}
  \caption{Time evolution of the GW amplitude $ h $ and maximum
    density $ \rho_\mathrm{max} $ for three representative models with
    different rotation profiles and initial rotation rates
    $ \beta_\mathrm{i} $.}
  \label{fig:generic_collapse_type}
\end{figure}


As in~\cite{ott_06_b}, in the entire investigated parameter space our
models yield GW burst signals of Type~I, i.e.\ the waveform exhibits a
positive pre-bounce rise and then a large negative peak, followed by a
ring-down (upper panel of Fig.~\ref{fig:generic_collapse_type}).
However, with respect to collapse dynamics and the relevant forces
halting the collapse, the models fall into two classes. While for
instance all models with the almost uniform rotation profile A1
experience a pressure-dominated bounce for which a Type~I waveform is
expected, models with profiles A2 or A3 \emph{and} sufficiently strong
initial rotation rate $ \beta_\mathrm{i} $ ($ \ge 4\% $ for A2;
$\ge 1.8\% $ for A3) show a \emph{single} centrifugal bounce at
subnuclear density (lower panel of
Fig.~\ref{fig:generic_collapse_type}).
Nevertheless, they also produce a Type~I waveform, as their core does
not re-expand after bounce to densities much less than bounce density
but immediately settles to a PNS after a short ring-down phase. What
obviously distinguishes models with pressure-dominated bounce from
those with centrifugal bounce is that the latter have GW signals
with significantly lower average frequencies. Note also that models
with very little rotation develop convective overturn of the
shock-heated layer immediately after shock stagnation (not to be
confused with the late-time convection discussed
in~\cite{mueller_04_a}), resulting in a lower-frequency contribution
to the post-bounce GW signal (see
Fig~\ref{fig:generic_collapse_type}).

In order to analyze the absence of Type~II signals, in particular for
cases with centrifugal bounce, we now separately investigate the
influence of GR, a microphysical EoS, and deleptonization on the
dynamics of rotational core collapse with different amounts and
distributions of angular momentum. When the pre-collapse iron core
starts to contract, its \emph{effective} adiabatic index
$ \gamma_\mathrm{eff} $ is lower than the critical value
$ \simeq 4 / 3 $ needed for stability against gravitational collapse.
Here $ \gamma_\mathrm{eff} $ is the sum of the adiabatic index
$ \gamma_\mathrm{eos} = \partial \ln P / \partial \ln \rho|_{Y_e, s} $
of the EoS (where $ P $ is the pressure, $ \rho $ the density, and
$ s $ the specific entropy of the fluid) and a possible correction
due to deleptonization (which can be significant until neutrino
trapping; see~\cite{moenchmeyer_91_a}). At this stage, both GR and
rotational effects (in our range of $ \beta_\mathrm{i} $) are
negligible in discussing stability. If the build-up of centrifugal
forces in the increasingly faster spinning core during collapse is
strong enough, contraction is halted and the core undergoes a
centrifugal bounce rather than reaching nuclear density (where EoS
stiffening would also stop the collapse).

A necessary condition for a centrifugal bounce at subnuclear densities
is that $ \gamma_\mathrm{eff} $ exceeds a critical rotation index
$ \gamma_\mathrm{rot} $. There exists a simple Newtonian analytic
relation, $ \gamma_\mathrm{rot} = (4 - 10 \beta_\mathrm{ic,b}) /
(3 - 6 \beta_\mathrm{ic,b}) $~\cite{tohline_84_a} (where
$ \beta_\mathrm{ic,b} $ is the inner core's rotation rate at bounce),
which works well in equilibrium, but is rather imprecise as a
criterion for centrifugal bounce in a dynamical situation. For
instance, for rotating core collapse models in Newtonian gravity with
a simple hybrid EoS~\cite{janka_93_a} and no deleptonization (where
$ \gamma_\mathrm{eff} = \gamma_\mathrm{eos} $), we find that for our
range of initial rotation rates and
$ 1.24 \le \gamma_\mathrm{eff} \le 1.332 $, the analytic relation
strongly underestimates the actual $ \gamma_\mathrm{rot} $ by up to
$ \sim 0.2 $ at high $ \beta_\mathrm{ic,b} $. Furthermore,
$ \beta_\mathrm{ic,b} $ is a result of the evolution, depending on the
initial parameters $ A $ and $ \beta_\mathrm{i} $ of the pre-collapse
core in an a-priori unknown way.

\begin{figure}
  \includegraphics[width = 8.5 cm]{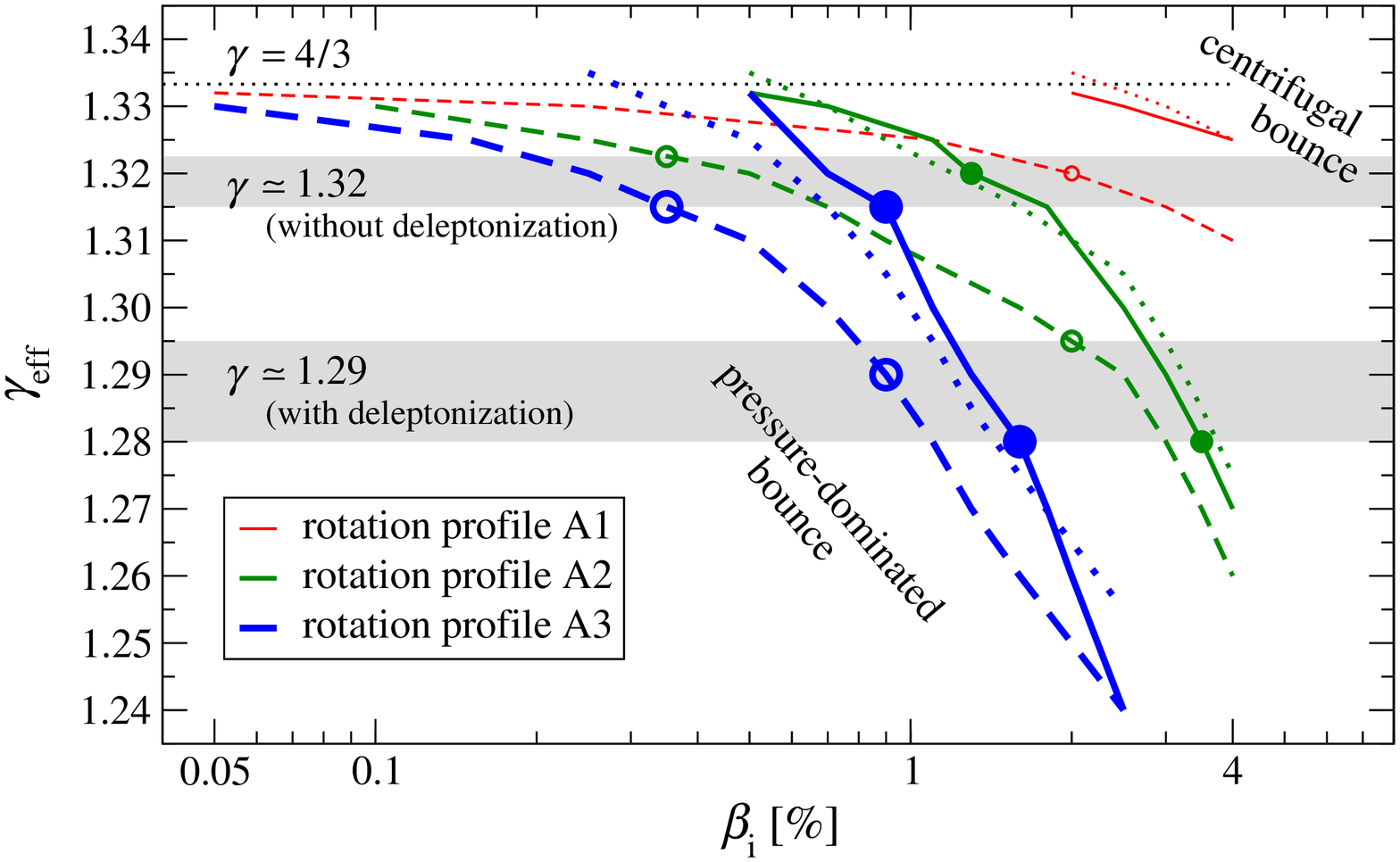}
  \caption{Boundary between pressure-dominated and centrifugal bounce
    in the $ \gamma_\mathrm{eff} $--$ \beta_\mathrm{i} $ plane for
    models using the hybrid EoS in Newtonian gravity (dashed lines) and
    GR (solid lines). Dotted lines show the Newtonian results
    shifted by $ - \Delta \gamma_\mathrm{gr} = 0.015 $. The transition
    points for models using the microphysical EoS without and with
    deleptonization, again for Newtonian gravity (circles) and GR
    (bullets), lie in the shaded areas around
    $ \gamma_\mathrm{eff} \simeq 1.32 $ and $ 1.29 $, respectively.}
  \label{fig:relativistic_corrections}
\end{figure}

For this reason, we determine the boundary between pressure-dominated
and centrifugal bounce in terms of the parameters $ \beta_\mathrm{i} $
and $ \gamma_\mathrm{eff} $ by systematic numerical simulations. For
models with a simple hybrid EoS~\cite{janka_93_a} and no
deleptonization, using the same initial density profile as in the
microphysical models, the results are shown in
Fig.~\ref{fig:relativistic_corrections}, both in the Newtonian case
(dashed lines) and in GR (solid lines). Apparently, for our choice of
initial rotation the influence of GR can be approximated by adding an
offset of $ - \Delta \gamma_\mathrm{gr} \simeq 0.015 $ to the
Newtonian results (dotted lines). This gives a quantitative measure of
the GR effects on rotational core collapse, which is in agreement
with~\cite{dimmelmeier_02_a}. Note that $ \Delta \gamma_\mathrm{gr} $
is negative because GR effectively acts like a softening of the EoS.

Fig.~\ref{fig:relativistic_corrections} also shows for each rotation
profile the locations where the transition between pressure-supported
and centrifugal bounce occurs when the microphysical EoS is used and
$ \beta_\mathrm{i} $ is gradually increased. These transitions are
marked on the different boundary lines and allow the identification of
the $ \gamma_\mathrm{eff} $ value where models with the hybrid EoS
make this transition. For simulations with microphysical EoS but no
deleptonization we find that in all cases the transition occurs near
$ \gamma_\mathrm{eff} \simeq 1.32 $ (highlighted by the upper grey
band in Fig.~\ref{fig:relativistic_corrections}). This value agrees
with the average of $ \gamma_\mathrm{eos} $ for the microphysical EoS
at densities between $ 10^{12} $ and $ 10^{14} \mathrm{\ g\ cm}^{-3} $,
which is the most relevant range for the collapse dynamics. Thus the
type of bounce obtained with the microphysical EoS is well reproduced
by a hybrid EoS with $ \gamma_\mathrm{eos} \simeq 1.32 $.

Deleptonization before neutrino trapping reduces
$ \gamma_\mathrm{eff} $ compared to $ \gamma_\mathrm{eos} $ locally
according to $ \Delta \gamma_e = \frac{4}{3} \, \delta \ln Y_e /
\delta \ln \rho < 0 $~\cite{van_riper_81_a}. Above trapping an
additional positive correction $ \Delta \gamma_\nu \approx
\delta (P_\nu / P) / \delta \ln \rho $ due to neutrino pressure
effects must be considered. From the $ Y_e $--$ \rho $ trajectories
used to describe the deleptonization during core collapse, we
estimate values between $ -0.03 $ and $ -0.02 $ for
$ \Delta \gamma_e + \Delta \gamma_\nu $, again in the density regime
relevant for the bounce dynamics. Therefore
$ \gamma_\mathrm{eff} \approx 1.29 $ is expected for models with
microphysical EoS \emph{and} deleptonization. Again this agrees with
the simulations. Fig.~\ref{fig:relativistic_corrections} shows that
the circles and bullets marking those models on the different boundary
lines (for the investigated initial rotation profiles with Newtonian
gravity and GR) all lie in the range of values indicated by the lower
grey band.

The finding that deleptonization decreases $ \gamma_\mathrm{eff} $ to
about $ 1.29 $ explains the absence of Type~II GW signals for our most
sophisticated models. When a hybrid EoS is used, the subgroup of such
models showing multiple centrifugal bounces and subsequent strong
re-expansion phases of the inner core occupies only a small area in
the $ \gamma_\mathrm{eff} $--$ \beta_\mathrm{i} $ plane. It is located
at $ \gamma_\mathrm{eff} \ge 1.31 $ for all of our initial rotation
states both in the Newtonian case and in GR, i.e.\ significantly above
the value of $ \gamma_\mathrm{eff} \simeq 1.29 $ that characterizes
the microphysical models if deleptonization is included.

The value $ \gamma_\mathrm{eff} = 1.29 $ captures well the
deleptonization effects in the density regime above neutrino trapping,
and therefore models with hybrid EoS and $ \gamma_\mathrm{eos} = 1.29 $
exhibit the same collapse and bounce behavior in this dynamical
phase. For $ \gamma_\mathrm{eos} = 1.29 $ the hybrid EoS corresponds
to an effective average $ Y_e $ of $ \approx 0.237 $, and thus for
models with this EoS we find a small mass
$ M_\mathrm{ic} \sim 0.1 \mbox{\,--\,} 0.3 \, M_\odot $ of the inner
core at bounce (higher for more rapid rotation), consistent with the
theory of self-similar collapse~\cite{yahil_83_a}. For models with
microphysics, however, $ \gamma_\mathrm{eff} $ is much closer to
$ 4 / 3 $ in the early collapse phase where the initial value of
$ M_\mathrm{ic} $ is determined. At intermediate densities below and
around neutrino trapping, deleptonization indeed reduces
$ M_\mathrm{ic} $, but only to about
$ 0.5 \mbox{\,--\,} 0.9 \, M_\odot $, which is in agreement with
recent spherically symmetric GR results using Boltzmann neutrino
transport~\cite{hix_03_a}. These large values of $ M_\mathrm{ic} $
explain the complete absence of Type~III GW burst signals in our
models (see also~\cite{kotake_03_a}).


The generic nature of the GW burst signal has several important
implications for prospective detectability. The limitation to a unique
Type~I waveform for a very broad range of rotation states of the
progenitor core will very likely facilitate the use of more powerful
and finetuned data analysis methods in GW detectors. To this end, we
offer our results in a waveform catalog~\cite{wave_catalog}.
Note that almost all investigated models result in a
pressure-dominated bounce and instantaneous formation of a PNS with
similar average density and compactness. For these models, whose
rotation rates at bounce span two orders of magnitude
($ 0.2\% \lesssim \beta_\mathrm{ic,b} \lesssim 20\% $), the maxima
$ f_\mathrm{max} $ of their waveform's frequency spectrum lie in a
very narrow range with an average of
$ \bar{f}_\mathrm{max} \simeq 718 \mathrm{\ Hz} $.

\begin{figure}
  \includegraphics[width = 8.5 cm]{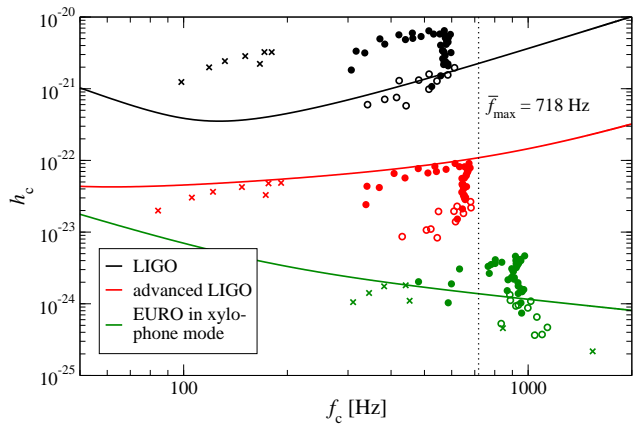}
  \caption{Location of the GW burst signals from the core bounce for
    all models in the $ h_\mathrm{c} $--$ f_\mathrm{c} $ plane
    relative to the sensitivity curves of various GW interferometer
    detectors (as color-coded). The sources are at a distance of
    $ 10 \mathrm{\ kpc} $ for LIGO, $ 0.8 \mathrm{\ Mpc} $ for
    advanced LIGO, and $ 15 \mathrm{\ Mpc} $ for EURO. Open circles
    denote models where the filtered-out early postbounce convection
    contributes significantly to the original signal, and crosses show
    models undergoing centrifugal bounce.}
  \label{fig:signal_detectability}
\end{figure}

This clustering in frequency is reflected in
Fig.~\ref{fig:signal_detectability}, where we plot the
(detector-dependent) frequency-integrated characteristic waveform
amplitude $ h_\mathrm{c} $ against the characteristic frequency
$ f_\mathrm{c} $ (Eq.~(31) in~\cite{thorne_87_a}) for all 54 GR models
with microphysical EoS and deleptonization. We assume optimal
orientation of source and detector, and in cases with
pressure-dominated bounce remove the lower-frequency contribution from
the post-bounce convective overturn by cutting the spectrum below
$ 250 \mathrm{\ Hz} $, as we are only interested in the GW signal from
the bounce and ring-down. Fig.~\ref{fig:signal_detectability} shows
that while current LIGO class interferometric detectors are only
sensitive to signals coming from an event in the Milky Way, advanced
LIGO could marginally detect some signals from other galaxies in the
Local Group like Andromeda. For the proposed EURO detector~\cite{euro}
in xylophone mode, we expect a very high signal-to-noise ratio (which
is $ h_\mathrm{c} $ divided by the detector sensitivity at
$ f_\mathrm{c} $). This detector could also measure many of the
computed signals at a distance of $ 15 \mathrm{\ Mpc} $, i.e.\ in the
Virgo cluster.

Note that detectability could be enhanced by (i) a network of
interferometers, (ii) the support by resonant detectors, which is
facilitated by the narrow range of $ f_\mathrm{max} $, and (iii) the
use of more powerful data analysis methods based on the waveforms'
similarity and robustness. A serious obstacle for detection is the low
event rate of $ \sim 1 \mathrm{\ yr}^{-1} $ within
$ 15 \mathrm{\ Mpc} $, which is further reduced by assuming that only
a fraction of all progenitors rotate fast enough to have a strong GW
bounce signal. Nevertheless, our results can serve as a guideline for
a possible frequency narrow-banding of future interferometers (which
can significantly boost sensitivity), as well as for choosing the
optimal configuration in the planning of resonant detectors. In
addition, as a core collapse may be accompanied by other GW emission
mechanisms like late-time convection (also in a nonrotating core) or
bar-mode instabilities, the \emph{total} GW signal strength and
duration could be significantly higher than predicted here.

As a downside, the generic properties of the GW burst signal introduce
a frequency degeneracy into the signal inversion problem.
Consequently, in the case of a detection it is difficult to extract
details about the rotation state of the pre-collapse core, because a
large part of the corresponding parameter space yields a
pressure-dominated bounce and thus signals with very similar values
for $ f_\mathrm{max} $ and (depending on the detector) also for
$ f_\mathrm{c} $. On the other hand, as $ f_\mathrm{max} $ directly
depends on the compressibility of the nuclear EoS at bounce, 
determining this frequency from the GW burst signal can help to
constrain the EoS properties around nuclear density.


We thank D.~Shoemaker for helpful discussions. This work was supported
by DFG (SFB/TR~7 and SFB~375).



\begin{thebibliography}{99}

\bibitem{mueller_82_a}
  E.~M\"uller,
  Astron. Astrophys. \textbf{114}, 53 (1982).

\bibitem{moenchmeyer_91_a}
  R.~M\"onchmeyer, G.~Sch\"afer, E.~M\"uller, and R.~Kates,
  Astron. Astrophys. \textbf{246}, 417 (1991).

\bibitem{zwerger_97_a}
  T.~Zwerger and E.~M\"uller,
  Astron. Astrophys. \textbf{320}, 209 (1997).

\bibitem{dimmelmeier_02_a}
  H.~Dimmelmeier, J.~Font, and E.~M\"uller,
  Astron. Astrophys. \textbf{393}, 523 (2002).

\bibitem{kotake_03_a}
  K.~Kotake, S.~Yamada, and K.~Sato,
  Phys. Rev. D \textbf{68}, 044023 (2003).

\bibitem{ott_04_a}
  C.~D. Ott, A.~Burrows, E.~Livne, and R.~Walder,
  Astrophys. J. \textbf{600}, 834 (2004).

\bibitem{ott_06_b}
  C.~D. Ott, H.~Dimmelmeier, A.~Marek, H.-T. Janka, I.~Hawke,
  B.~Zink, and E.~Schnetter,
  Phys. Rev. Lett. (2006), submitted.

\bibitem{rampp_98_a}
  M.~Rampp, E.~M\"uller, and M.~Ruffert,
  Astron. Astrophys. \textbf{332}, 969 (1998).

\bibitem{mueller_04_a}
  E.~M\"uller, M.~Rampp, R.~Buras, H.-T. Janka, and D.~H. Shoemaker,
  Astrophys. J. \textbf{603}, 221 (2004).

\bibitem{shibata_05_a}
  M.~Shibata and Y.-I. Sekiguchi,
  Phys. Rev. D \textbf{71}, 024014 (2005).

\bibitem{ott_06_a}
  C.~D. Ott, A.~Burrows, L.~Dessart, and E.~Livne,
  Phys. Rev. Lett. \textbf{96}, 201102 (2006).

\bibitem{dimmelmeier_05_a}
  H.~Dimmelmeier, J.~Novak, J.~A. Font, J.~M. Ib\'a\~nez, and E.~M\"uller,
  Phys. Rev. D \textbf{71}, 064023 (2005).

\bibitem{shibata_04_a}
  M.~Shibata and Y.-I. Sekiguchi,
  Phys. Rev. D \textbf{69}, 084024 (2004).

\bibitem{shen_98_a}
  H.~Shen, H.~Toki, K.~Oyamatsu, and K.~Sumiyoshi,
  Prog. Theor. Phys. \textbf{100}, 1013 (1998).

\bibitem{marek_05_a}
  A.~Marek, H.-T. Janka, R.~Buras, M.~Liebend\"orfer, and M.~Rampp,
  Astron. Astrophys. \textbf{443}, 201 (2005).

\bibitem{liebendoerfer_05_a}
  M.~Liebend\"orfer,
  Astrophys. J. \textbf{633}, 1042 (2005).

\bibitem{langanke_00_a}
  K.~Langanke and G.~Mart\'{\i}nez-Pinedo,
  Nucl. Phys. A \textbf{673}, 481 (2000).

\bibitem{woosley_02_a}
  S.~E. Woosley, A.~Heger, and T.~A. Weaver,
  Rev. Mod. Phys. \textbf{74}, 1015 (2002).

\bibitem{tohline_84_a}
  J.~E. Tohline,
  Astrophys. J. \textbf{285}, 721 (1984).

\bibitem{janka_93_a}
  H.-T. Janka, T.~Zwerger, and R.~M\"onchmeyer,
  Astron. Astrophys. \textbf{268}, 360 (1993).

\bibitem{van_riper_81_a}
  K.~A. van Riper and J.~M. Lattimer,
  Astrophys. J. \textbf{249}, 270 (1981).

\bibitem{yahil_83_a}
  A.~Yahil, 
  Astrophys. J. \textbf{265}, 1047 (1983).
  
\bibitem{hix_03_a}
  W.~R. Hix, O.~E.~B. Messer, A.~Mezzacappa, M.~Lieben\-d\"orfer,
  J.~Sampaio, K.~Langanke, D.~J. Dean, and G.~Mart\'{\i}nez-Pinedo,
  Phys. Rev. Lett. \textbf{91}, 201102 (2003).

\bibitem{wave_catalog}
  www.mpa-garching.mpg.de/rel\_hydro/wave\_catalog.shtml.
  
\bibitem{thorne_87_a}
  K.~S. Thorne, in
  \emph{300 Years of Gravitation},
  edited by S.~W. Hawking and W.~Israel
  Cambridge University Press, Cambridge, UK, 1987), p.~369.
  
\bibitem{euro}
  www.astro.cardiff.ac.uk/geo/euro/.

\end{thebibliography}
\end{document}